\documentclass[%
reprint,
runinaddress,
nofootinbib,
 amsmath,amssymb,
 aps,
]{revtex4-1}
\usepackage{lipsum}
\usepackage{graphicx}
\usepackage{dcolumn}
\usepackage{bm}
\usepackage{color}
\usepackage{tikz}
\usetikzlibrary{matrix}
\definecolor{awesome}{rgb}{1.0, 0.13, 0.32}
\definecolor{azure(colorwheel)}{rgb}{0.0, 0.5, 1.0}
\definecolor{guppiegreen}{rgb}{0.0, 1.0, 0.5}
\usepackage{hyperref}

\hypersetup{
    bookmarks=true,         
    unicode=false,          
    pdftoolbar=true,        
    pdfmenubar=true,        
    pdffitwindow=false,     
    pdfstartview={FitH},    
    pdftitle={Exact instantons via worldline deformations},    
    pdfauthor={A. Akal},     
    pdfnewwindow=true,      
    colorlinks=true,       
    linkcolor=azure(colorwheel),          
    citecolor=awesome,        
    filecolor=azure(colorwheel),      
    urlcolor=guppiegreen           
}

\newcommand{\lc}{\left(}
\newcommand{\rc}{\right)}
\newcommand{\lC}{\left[}
\newcommand{\rC}{\right]}
\newcommand{\eqn}[1]{\begin{align}#1\end{align}}
\newcommand{\eqnsplit}[1]{\begin{align}\begin{split}#1\end{split}\end{align}}
\newcommand*{\bfrac}[2]{\genfrac{}{}{0pt}{}{#1}{#2}}
\begin{document}
\title{Exact instantons via worldline deformations}
\author{A. Akal}

\affiliation{Theory Group\\
DESY\\
D-22607 Hamburg, Germany\\}

\date{\today}

\begin{abstract}
The imaginary part of the one loop effective action in external backgrounds
can be efficiently computed using worldline instantons which are closed periodic paths in spacetime.
Exact solutions for nonstatic backgrounds are only known in certain cases. In this paper, we propose a novel technique
allowing the construction of
further exactly solvable models.
In order to do so, we introduce a deformation function which maps the worldline instantons for a given model to the
closed periodic stationary paths of a new model.
Executing this procedure iteratively results in a chain of infinitely many solvable models.
Similar ideas were applied to topological and nontopological defects in quantum field theory.
We explicitly
discuss the tunneling exponential in the Schwinger pair creation rate and illustrate
the validity of the proposed technique for well-known cases.
\end{abstract}
\maketitle

\section{Introduction}
\label{Sec:intro}
Pair creation from the vacuum in the presence of macroscopic gauge fields is one of the most intriguing nonperturbative results in quantum field theory with a clear physical prediction \cite{Schwinger:1951nm}. 

However, this nowadays called Schwinger effect has so far eluded a direct experimental verification.
In the weak field and weak coupling limit, the creation rate
is determined by the imaginary part of the one loop Euler-Heisenberg effective action.
In the simplest scenario, i.e. when the system is subjected to a static electric background (standard Schwinger mechanism),
the pair creation rate $\mathcal R$ \cite{Cohen:2008wz}
\eqn{
\mathcal{R} \simeq \frac{E^2}{(2 \pi)^{3}} e^{-\mathcal{W}_0},\qquad \mathcal{W}_0 = \pi \frac{E_\mathrm{S}}{E}
\label{eq:static-rate}
}
is dramatically suppressed for field strengths $E$ much below
the critical value $E_\mathrm{S} = m^2$, where $m$ denotes the mass of the particles.
Notably, the nonanalytic dependence $\mathcal W_0 \sim 1/E$ in the exponent clearly manifests the characteristic nonperturbative behavior.
In the case of nonstatic Abelian backgrounds, it has been shown that $\mathcal R$ is very sensitive to the background shape. Generally,
large background inhomogeneities
tend to trigger an enormous enhancement even for $E \ll E_\text S$ \cite{Brezin:1970xf,Dunne:2005sx,Schutzhold:2008pz}.
Backgrounds of this type may, for instance, be realized with upcoming lasers \cite{Bulanov:2010ei}.

Beyond the static limit, preferably for purely electric backgrounds, investigations mostly rely on numerical
computation techniques, e.g. \cite{Gies:2005bz,Dunne:2006ur,Hebenstreit:2011wk,Akal:2014eua,Otto:2015gla}. Analytic investigations have so far been performed only
for certain configurations, e.g. \cite{Brezin:1970xf,Nikishov:1970br,Kim:2000un,Kim:2003qp,Schutzhold:2008pz,Popov:1971ff,KeskiVakkuri:1996gn,Dunne:2005sx,Kleinert:2008sj,bai2012electron,Ilderton:2015qda,Adorno:2015ibo}.
Nevertheless, a deeper understanding of the impact
of more complicated backgrounds is necessary to better understand the
formation process, especially, to unveil interesting insights about critical points \cite{Akal:2017ilh}, nonlocal effects \cite{Dunne:2005sx} and also universal features \cite{Pimentel:2018nkl}. Technically, this task is highly challenging and cannot be
pursued with conventional field theoretic methods.

A powerful technique relies on the first-quantized string inspired formulation \cite{Bern:1991aq}. In particular,
the semiclassical treatment via worldline
instantons proves highly efficient, see e.g. \cite{Affleck:1981bma,Dunne:2005sx,Dunne:2006ur,bai2012electron,Gordon:2014aba,Medina:2015qzc,Brown:2015kgj,Gould:2017fve,Akal:2017ilh,Akal:2017sbs}
and also \cite{Silverstein:1996xp,Dietrich:2014ala,Basar:2015xna,Dumlu:2017kfp} for applications in other contexts.
Within this approach,
the imaginary part
of the effective action just becomes determined by closed periodic paths
in spacetime.
Principally, the basic
task then is to find the associated worldline instantons. There exist only a few cases in the literature for which the stationary worldline paths have been computed exactly, see e.g. \cite{Dunne:2005sx,Ilderton:2015qda}.
Even the numerical treatment is highly nontrivial due to the underlying nonlinear equations \cite{Dunne:2006ur,Akal:2017sbs}.

In this paper, we propose a novel technique which enables the construction of arbitrary many solvable models
for which the corresponding worldline instantons can be obtained by
deforming a known solution according to a particularly constructed map.\footnote{These classical worldline solutions determine the damping factor $\mathcal W_0$ occurring in the pair production rate $\mathcal R$ as in \eqref{eq:static-rate}.}
Similar techniques have been utilized to generate solvable models giving rise to topological and nontopological defects in quantum field theory \cite{Bazeia:2002xg,Bazeia:2005hu}.\footnote{It shall also be emphasized that making use of diffeomorphisms in order to generate new solutions to the underlying equations of motion is a well-established technique in the context of general relativity.}

The remaining part of this paper is structured as follows:
in section \ref{Sec:wis}, we introduce the Schwinger pair creation rate and
the corresponding instanton equations.
In section \ref{Sec:deformed-defects}, we recap
the deformation procedure in the context of defects in field theory.
In sections \ref{Sec:recast} and \ref{Sec:deformed-worldlines}, being the main parts of this paper, we first reshape the underlying saddle point equations and propose the deformation procedure for worldline instantons.
In section \ref{Sec:discuss}, we discuss our results and validate our predictions
for certain solvable models.
We finish with a brief summary and outlook in section \ref{Sec:summ}.

\section{Worldline instantons}
\label{Sec:wis}
We start with a brief discussion of vacuum decay in the presence of an external Abelian background.
The general expression for the decay probability $\mathcal P$ against the creation of matter-antimatter pairs is \cite{Schwinger:1951nm}
\eqn{
\mathcal P \simeq 1 - e^{- 2 \mathcal R}.
}
As usual, we neglect the contribution coming from the dynamical part of the vector field. Then one can rewrite the logarithm of the kinetic operator determinant, here given in scalar quantum electrodynamics,
and apply the quantum mechanical path integral representation
for the resulting trace using position eigenstates.
The outcome, after analytically continuing to Euclidean spacetime, is the following worldline representation
\begin{align}
\mathcal{R} \simeq  \int_0^\infty \frac{ds}{s}\ e^{ -m^2 s } \oint \mathcal{D}x\ e^{ -\int_0^s d\tau\ \lC \frac{\dot x^2(\tau)}{4} + i \mathcal{A} \cdot \dot x(\tau)\rC },
\label{eq:rate-weak}
\end{align}
where $\mathcal{D}x$ denotes the path integral measure over the closed worldlines with periodic boundaries,
$\mathcal A^\mu$ is the external vector potential and $s$ denotes the proper time.

Performing
a saddle point analysis in $s$ and
expanding the resulting worldline action\footnote{More specifically, starting from the Euclidean worldine representation in \eqref{eq:rate-weak}, one first performs a variable rescaling $\tau \rightarrow u s$, where $u \in [0,1]$, and uses the so-called large mass approximation $m \sqrt{\int_0^1 du\ \dot{x}^2 } \gg 1$, which, for instance, in Schwinger's original static field scenario corresponds to the weak field limit $E_\text{S}/E \gg 1$.}
\begin{align}
\mathcal{W} \simeq m
a
+ i \int_0^1 du\ \mathcal{A} \cdot \dot x(u)
\label{eq:wl-action-weak}
\end{align}
in the fluctuations over the saddle points, called worldline instantons,\footnote{These are the classical worldline solutions to the Euclidean stationary equations and, hence, share parallels with ordinary instantons in quantum field theory, thus the naming.} obeying the boundary condition $x_\mu(0) = x_\mu(1)$,
yields
\begin{align}
\mathcal{R} \simeq \frac{1}{\sqrt{\text{Det} M}} e^{ - \mathcal{W}_0},\qquad \mathcal{W}_0 \equiv \mathcal{W} \left[\mathrm{instanton}\right],
\label{eq:R-tot}
\end{align}
where $M_{\mu \nu}$ is the second order variation operator and the constant $a^2 \equiv \dot x^2$ is taken to be an internal invariant constant on the worldline, i.e.~$\partial_u a = 0$ \cite{Affleck:1981bma}.
The expression in front of the exponential leads to the quantum fluctuation prefactor.
For the present discussion we only focus on the tunneling exponential involving the damping factor $\mathcal W_0$.

Minimizing the resulting exponent
shows that
the closed periodic instanton path is determined by the following set of equations
\eqn{
m \ddot x^\mu &= i a \mathcal F^{\mu \nu} \dot x^\nu.
\label{eq:instanton}
}

For simplifying reasons, here, we consider time dependent backgrounds with the general form
\eqn{
\mathcal A_3 = -i E F(x_4)
\label{eq:A3}
}
where $F$ is assumed to be an odd function in $x_4$.\footnote{We consider the spacetime to be two-dimensional, where the Euclidean components $x_3$ and $x_4$ denote the corresponding spatial and temporal directions, respectively.}

As will be seen in section~\ref{Sec:recast}, such backgrounds allow a straightforward rewriting of the underlying stationary equations which resemble the one occurring in the context of defects in field theory.
After inserting $\mathcal A_3$ into the instanton equations from \eqref{eq:instanton}, we end up with the
following system of coupled differential equations
\eqnsplit{
\ddot x_3 &= - \frac{a E}{m} F'(x_4) \dot x_4,\\
\ddot x_4 &= + \frac{a E}{m} F'(x_4) \dot x_3,
\label{eq:ddot-x3-x4}
}
where $\dot{x}_{3,4} \equiv \partial_u x_{3,4}$ and $F'(x_4)  \equiv \partial_{x_4} F(x_4)$.
Evaluating the worldline action $\mathcal W$ on the solution of the coupled differential equations in
\eqref{eq:ddot-x3-x4} gives the exponential part in $\mathcal R$.
In what follows, we briefly comment on some ideas regarding defect solutions in field theory.

\section{Deforming defects}
\label{Sec:deformed-defects}
Defects are important in high energy \cite{vilenkin2000cosmic,manton2004topological} as well as condensed matter physics \cite{chaikin1995principles}, e.g. in the context of gravity with warped geometry involving extra dimensions, and gap opening in the mass spectrum of charge carriers, respectively.
Defects can be of topological and nontopological type.
Quantum field theories which support spontaneous symmetry breaking usually give rise to the former kind. Examples are kinks, domain walls and monopoles to name a few.

Models in terms of real scalar fields
usually have kinklike or lumplike defects \cite{Gani:2014gxa,Avelar:2007hx}.
Interestingly, in a spatially one-dimensional solvable model of such kind, a particularly realized deformation of defects
can lead to new models with defects expressed in terms of the original defect solutions \cite{Bazeia:2002xg}, e.g., see also \cite{Granado:2020hmh}.
For instance, consider a scalar field theory in $1+1$ dimensions where fields and coordinates are taken to be dimensionless. Let its Lagrangian be
\eqn{
\mathcal L_\chi = \frac{1}{2} \partial_\mu \chi \partial^\mu \chi - V_\chi (\chi)
}
where $V_\chi$ is the specific potential in this model that we call the $\chi$ model.
The corresponding equations of motion determining the associated static solution are
\eqnsplit{
 \frac{d^2 \chi}{du^2} &= \frac{dV_\chi(\chi)}{d\chi},\\
 \lC \frac{d\chi}{du} \rC^2 &= 2 V_\chi(\chi).
 \label{eq:chi-th}
}
For the second line, it has been assumed that $\chi$
satisfies the following boundary conditions
\eqn{
\chi \rightarrow \bar \chi, \qquad \dot{\chi} \rightarrow 0\qquad (u \rightarrow -\infty ),
}
where $\dot{\chi} \equiv \partial_u \chi$.
Here, $\bar \chi$ denotes the critical point of the potential, i.e. $V'_\chi(\bar \chi) = 0$, where
$V_\chi(\bar \chi) = 0$.

Let us now suppose a second model described by the Lagrangian
\eqn{
\mathcal L_\phi = \frac{1}{2} \partial_\mu \phi \partial^\mu \phi - V_\phi (\phi)
}
for which the static solution is determined by
\eqnsplit{
 \frac{d^2 \phi}{du^2} &= \frac{dV_\phi(\phi)}{d\phi},\\
 \lC \frac{d\phi}{du} \rC^2 &= 2 V_\phi(\phi),
\label{eq:phi-th}
}
similar to before, but now depending on a different potential, i.e. $V_\phi \neq V_\chi$.

Assume that the former $\chi$ model described by $\mathcal L_\chi$ is exactly solvable and supports
finite energy defects. Then, we can
write the potential $V_\phi$ for the $\phi$ model in the following specific form \cite{Bazeia:2002xg}
\eqn{
V_\phi(\phi) = \frac{V_\chi(\chi \rightarrow f(\phi))}{\lC df/d\phi \rC^2}
\label{eq:Vphi-defects}
}
where $f$ is taken to be a real bijective function with nonzero first order derivative.
This particular deformation has the interesting consequence that the solution in the deformed $\phi$ model described by $\mathcal L_\phi$
is exactly given by
\eqn{
\phi = f^{-1}(\chi)
\label{eq:phi}
}
where $f^{-1}$ denotes the inverse of the deforming map and $\chi$ exactly solves the stationary equations in the original theory.
\section{Recasting the saddle point equations}
\label{Sec:recast}
Treating
worldline instantons in an analogous way as described above, we aim to
a work out a relation similar to \eqref{eq:Vphi-defects}, but now we want to relate two distinct models with different stationary worldline solutions.
Note that in this case, we deal with worldline components instead of a real quantum field as in section \ref{Sec:deformed-defects}. Nevertheless, we will show below that a close similarity is indeed present.
To start with, we first need to bring the corresponding
stationary equations into an appropriate shape.
We integrate the first line in the saddle point equations \eqref{eq:ddot-x3-x4} to obtain
\eqn{
\dot x_3 = - \frac{aE}{m} F(x_4).
}
The latter we then insert into the second expression which leads to the relation,
\eqn{
\ddot x_4 = - \frac{d}{dx_4} \lC \frac{aE}{\sqrt{2} m} F(x_4) \rC^2.
\label{eq:ddot-x4-v2}
}
It is advantageous to define a potential of the form
\eqn{
V(x_4) \equiv - \lC \frac{aE}{\sqrt{2} m} F(x_4) \rC^2
\label{eq:defV}
}
in order to write down
\eqn{
\dot x_4^2 = a^2 + 2 V(x_4)
}
which follows from the second line in \eqref{eq:ddot-x3-x4}.

Expressing the right-hand side of \eqref{eq:ddot-x3-x4} in terms of the potential in \eqref{eq:defV}, we can reformulate the corresponding stationary equations in the following way
\eqnsplit{
 \frac{d^2 x_4}{du^2} &= \frac{dV(x_4)}{dx_4},\\
 \lC \frac{dx_4}{du} \rC^2 &= a^2 + 2 V(x_4).
\label{eq:recast}
}
Note that the latter are quite similar to the equations in \eqref{eq:chi-th} and \eqref{eq:phi-th}, respectively,
with the occurrence of an additional
constant term
in the second equation being the only difference.
Here, we only treat the Euclidean time coordinate.
The spatial one can be computed from the latter.

\section{Worldline deformations}
\label{Sec:deformed-worldlines}
The similarity between the stationary equations \eqref{eq:chi-th} and \eqref{eq:recast}
encourages one to proceed as in section \ref{Sec:deformed-defects}. In order to stick to the same convention, we
first rename
the time coordinate,
\eqn{
x_{4} \rightarrow \chi.
}
Then, assuming a solvable model is given, we first write down the equations for its stationary solution
\eqnsplit{
 \frac{d^2 \chi}{du^2} &= \frac{dV_\chi(\chi)}{d\chi},\\
 \lC \frac{d\chi}{du} \rC^2 &= a_\chi^2 + 2 V_\chi(\chi)
 \label{eq:chi-wi}
}
where $a_\chi^2 = \dot\chi^2 + \dot x_3^2$ is the corresponding invariant.

Similarly, we consider a second model depending on some other potential $V_\phi$ for which we set
\eqn{
x_{4} \rightarrow \phi.
}
The associated stationary solution is determined by the equations
\eqnsplit{
 \frac{d^2 \phi}{du^2} &= \frac{dV_\phi(\phi)}{d\phi},\\
 \lC \frac{d\phi}{du} \rC^2 &= a_\phi^2 + 2 V_\phi(\phi)
\label{eq:phi-wi}
}
with the invariant $a_\phi^2 = \dot\phi^2 + \dot x_3^2$.
The main observation is that if we impose a deformation prescription of the form
\eqn{
a_\phi^2 + 2V_\phi(\phi) = \frac{ a_\chi^2 + 2V_\chi(\chi \rightarrow f(\phi)) }{ \lC df/d\phi \rC^2 }
\label{eq:mod-def}
}
then the stationary solution in the $\phi$ model in \eqref{eq:phi-wi} can be entirely expressed in terms of $\chi$  determined by \eqref{eq:chi-wi},
as in \eqref{eq:phi}.

In other words, if we know the exact worldline instantons in the $\chi$ model, then
the worldline instantons in the deformed $\phi$ model with potential $V_\phi$ that is constructed according to the prescription in \eqref{eq:mod-def},
where $f$ is an appropriately chosen deformation function,
are also known.

In order to show
that $\phi = f^{-1}(\chi)$, see \eqref{eq:phi}, is indeed the exact deformed solution,
we start by expressing the potential $V_\phi$ in terms of $a_\chi,a_\phi,V_\chi$ and $f$ using the relation
\eqref{eq:mod-def}, in the first line of \eqref{eq:phi-wi}. The resulting right-hand side then reads
\eqn{
\frac{d^2 \phi}{du^2} =
\frac{V_\chi'(f(\phi))}{f'(\phi)} - \frac{f''(\phi)}{[f'(\phi)]^3} \lC a_\chi^2 + 2 V_\chi(f(\phi)) \rC,
\label{eq:rhs1}
}
where we have used
$\partial_u a_\phi = 0$.
On the other hand, the direct computation of the second derivative via \eqref{eq:phi}
yields
\eqn{
\frac{d^2 \phi}{du^2} =  \frac{1}{f'(\phi)} \lC \frac{d^2 \chi}{du^2} \rC - \frac{f''(\phi)}{[f'(\phi)]^3} \lC \frac{d \chi}{du} \rC^2.
}
Replacing the first bracket by the first line in \eqref{eq:chi-wi} and the second bracket by the second line in \eqref{eq:chi-wi}, we immediately end up with equation \eqref{eq:rhs1} after setting $\chi = f(\phi)$. 

Hence, a deformation of the form \eqref{eq:mod-def} indeed turns out to be compatible with the exact solutions to the stationary equations \eqref{eq:chi-wi} and \eqref{eq:phi-wi} associated with the initially chosen undeformed and deformed potential, respectively, where $\phi$ is expressed as in \eqref{eq:phi}.

\section{Discussion}
\label{Sec:discuss}
In this section, the deformation technique is
discussed for certain examples.
\subsection{General aspects}
For simplifying reasons, we first introduce the following definitions
\eqn{
\tilde F(\chi) \equiv \omega F(\chi),\qquad \gamma \equiv \frac{m \omega}{E}
\label{eq:defs}
}
where $\omega$ denotes the frequency.
Here, $\gamma$ (usually referred to as the Keldysh parameter) is a useful dimensionless parameter regulating the background inhomogeneity.
With these definitions, the potential introduced in \eqref{eq:defV} can be rewritten as
\eqn{
V(\chi) = - \frac{a^2}{2 \gamma^2} \tilde F^2(\chi).
}
Then, after plugging the latter into \eqref{eq:mod-def},
we may write down the deformation function in terms of $\tilde F_\chi$, $\tilde F_\phi$ and the internal invariants $a_\chi$, $a_\phi$ leading to the expression
\eqn{
\lC f'(\phi) \rC^2 =
\lC \frac{a_\chi}{a_\phi} \rC^2 \lC \frac{\gamma ^2-\tilde F_\chi^2(f(\phi))}{\gamma ^2-\tilde F_\phi^2(\phi)} \rC.
}
Similarly, one can formulate the corresponding equation
for the inverse deformation function,
\eqn{
\lC (f^{-1})'(\chi) \rC^{-2} =
\lC \frac{a_\chi}{a_\phi} \rC^2 \lC \frac{\gamma ^2-\tilde F_\chi^2(\chi)}{\gamma ^2-\tilde F_\phi^2(f^{-1}(\chi))} \rC.
}
The formal solutions for the latter equations are of the following form
\begin{widetext}
\eqn{
f(\phi) &=
\lc \int^\chi \frac{ d\chi'}{\sqrt{ \gamma^2 - \tilde F_\chi^2(\chi')} }  \rc^{-1}
\lc \lC \frac{a_\chi}{a_\phi} \rC \lC \int^\phi \frac{ d\chi'}{\sqrt{ \gamma^2 - \tilde F_\phi^2(\chi')} } \rC \rc,\\
f^{-1}(\chi) &=
\lc \int^\phi \frac{ d\phi'}{\sqrt{ \gamma^2 - \tilde F_\phi^2(\phi')} }  \rc^{-1}
\lc \lC \frac{a_\phi}{a_\chi} \rC \lC \int^\chi \frac{ d\phi'}{\sqrt{ \gamma^2 - \tilde F_\chi^2(\phi')} } \rC \rc.
\label{eq:formal-sols}
}
\end{widetext}
As can be clearly seen, one has to compute the inverse of some integral function with the general form
\eqn{
H(x) \equiv \int^x h(u)\ du.
}
The latter
can be also written
in terms of the integrand. Namely, using basic arguments, it can be shown that the first order derivative of the inverse has to satisfy the following differential equation
\eqn{
(H^{-1})'(x) = 1/\lC h(H^{-1}(x)) \rC.
\label{eq:H-h}
}
However, in this way the direct computation of the deformation function $f$ and its inverse $f^{-1}$ for some given potentials $V_\chi$ and $V_\phi$ is
still difficult to achieve.

Note that the deformation technique
substantially simplifies the problem of finding
a suitable model for which the worldline instantons are known exactly. According to the deformation technique, we do not need to compute the underlying integrals in \eqref{eq:formal-sols} for that purpose. Instead, one may start with an appropriately chosen
deformation function in order to construct a suitable potential $V_\phi$ by differentiating the deformation function. This procedure is substantially easier to
undertake. The difference between both strategies is sketched in figure \ref{fig:difference} below.
\begin{figure}[h!]
\centering
\begin{tikzpicture}
  \matrix (m) [matrix of math nodes,row sep=3em,column sep=4em,minimum width=2em]
  {
     V_\phi & f\\
     f & V_\phi\\
  };
  \path[-stealth]
    (m-1-1) edge node [above] {$\int \frac{ d\phi}{\sqrt{ \gamma^2 - \tilde F_\phi^2(\phi)} }$} (m-1-2)
    (m-2-1) edge node [above] {$\partial_\phi f$} (m-2-2);
\end{tikzpicture}
\caption{Deformation technique (below) compared to the standard route (top). The task of computing a complicated integral is replaced by the computation of a first order derivative.}
\label{fig:difference}
\end{figure}
Taking that into account,
the stationary worldline action for the deformed model depending on the constructed potential $V_\phi$
can be compactly written as
\eqn{
\mathcal W_0 = ma_\phi \lC 1 + \frac{1}{\gamma^2} \int_0^1 du\ \tilde F_\phi^2 \lc f^{-1}(\chi) \rc \rC.
}
\subsection{Constructing exactly solvable models}
The deformation strategy can be repeated once a new solvable model has been constructed.
It is therefore possible to generate many other exactly solvable models. The required steps can be specified in the following order:
\begin{enumerate}
 \item Choose an appropriate deformation function $f$ and compute its inverse $f^{-1}$.
 \item Choose an exactly solvable model depending on some potential $V_\chi$.
 \item Following the prescription of \eqref{eq:mod-def}, deform the initial solvable model to construct a
 new model with potential $V_\phi$.
 \item The exact instanton solution for the new model is then given by
 $\phi = f^{-1}(\chi)$.
 \item Repeat the latter steps (2 - 4) to construct a chain
of arbitrary many exactly solvable models.
\end{enumerate}
A separate chain of solvable models is possible if one operates with another suitable deformation function $\tilde f \neq f$ as sketched in figure \ref{fig:chains}. Here, one should notice that $f$ principally has to be normalized (in each iteration step), since the underlying quantities are in general not dimensionless.
This is necessary to generate a reasonable background.
\begin{figure}[h!]
\centering
\begin{tikzpicture}
  \matrix (m) [matrix of math nodes,row sep=3em,column sep=4em,minimum width=2em]
  {
     \bfrac{V_\chi}{\chi} & \bfrac{V_{\phi}}{\phi} & \bfrac{V'_{\phi'}}{\phi'} & \cdots \\
     \bfrac{V_{\chi}}{\chi} & \bfrac{\tilde V_{\tilde\phi}}{\tilde\phi} & \bfrac{\tilde V'_{\tilde\phi'}}{\tilde \phi'} & \cdots \\
  };
  \path[-stealth]
    (m-1-1) edge node [above] {$f$} (m-1-2)
            edge node [below] {$f^{-1}$} (m-1-2)
    (m-1-2) edge node [above] {$f$} (m-1-3)
	    edge node [below] {$f^{-1}$} (m-1-3)
    (m-1-3) edge node [above] {$f$} (m-1-4)
	    edge node [below] {$f^{-1}$} (m-1-4)
    (m-2-1) edge node [above] {$\tilde f$} (m-2-2)
            edge node [below] {$\tilde f^{-1}$} (m-2-2)
    (m-2-2) edge node [above] {$\tilde f$} (m-2-3)
	    edge node [below] {$\tilde f^{-1}$} (m-2-3)
    (m-2-3) edge node [above] {$\tilde f$} (m-2-4)
	    edge node [below] {$\tilde f^{-1}$} (m-2-4)
    (m-1-1) edge [dotted,-] (m-2-1)
	    -- node {$=$} (m-2-1)
    (m-1-2) edge [dotted,-] (m-2-2)
	    -- node {$\neq$} (m-2-2)
    (m-1-3) edge [dotted,-] (m-2-3)
	    -- node {$\neq$} (m-2-3)
            ;
\end{tikzpicture}
\caption{Two separate chains of exactly solvable models generated by applying two distinct deformation functions $f$ and $\tilde f$, respectively.}
\label{fig:chains}
\end{figure}
In the following, we discuss certain well-known solvable models to
confirm the validity of the proposed prescription \eqref{eq:mod-def}. We choose both models
so that $f$ as well as $f^{-1}$ can be computed directly. This we require in order to test the predicted solution of the form \eqref{eq:phi}.
To do so, we assume a static electric background described by $\mathbf{E} = E \hat x_3$ for the undeformed $\chi$ model. The corresponding potential function reads
\eqn{
\tilde F_\chi (\chi) = \omega \chi.
\label{eq:Ftilde-stat}
}
As mentioned previously, this is the simplest scenario and was first considered by Affleck et al. within the semiclassical worldline approach. The stationary solutions are the well known circular worldline instantons
determined by \cite{Affleck:1981bma}
\eqnsplit{
\chi &= (\gamma/\omega) \sin(2 \pi n u),\\
a_\chi &= (\gamma/\omega) 2 \pi n.
\label{eq:circular-wi}
}
Here, $n \in \mathbb N$ is usually called the winding number where integers $n > 1$ correspond to higher order instanton contributions which can be understood as the coherent creation of $n$ pairs \cite{Lebedev:1985bj}.

\subsection{A trivial example}
In the following, we compute $f$ directly. As noted before, the main advantage of the deformation technique is basically the opposite direction, means choosing $f$ and constructing $V_\phi$ afterwards. However, for illustrative reasons, we choose certain models having exact solutions. A trivial example is given when
we assume a static background for the deformed model, i.e.
\eqn{
\tilde F_\phi (\phi) = \omega \phi.
\label{eq:Ftilde-stat2}
}
Then, plugging the potential functions \eqref{eq:Ftilde-stat} and \eqref{eq:Ftilde-stat2} into \eqref{eq:mod-def} results in the following deformation function
\eqn{
f(\phi) =
\frac{\gamma \tan\lc \frac{a_\chi}{a_\phi} \tan^{-1}\lc \frac{\omega \phi }{\sqrt{\gamma^2 - (\omega \phi )^2}} \rc \rc}{\omega \sqrt{ 1 + \tan^2\lc \frac{a_\chi}{a_\phi} \tan^{-1}\lc \frac{\omega \phi }{\sqrt{\gamma^2 - (\omega \phi )^2}} \rc \rc }}.
}
Its inverse reads
\eqn{
f^{-1}(\chi) =
\frac{\gamma \tan\lc \frac{a_\phi}{a_\chi} \tan^{-1}\lc \frac{\omega \chi }{\sqrt{\gamma^2 - (\omega \chi )^2}} \rc \rc}{\omega \sqrt{ 1 + \tan^2\lc \frac{a_\phi}{a_\chi} \tan^{-1}\lc \frac{\omega \chi }{\sqrt{\gamma^2 - (\omega \chi )^2}} \rc \rc }}.
}
It is straightforward to verify that inserting the expressions from \eqref{eq:circular-wi} into $f^{-1}$ yields
the expected circular worldline instanton determined by
\eqnsplit{
\phi &= (\gamma/\omega) \sin(2 \pi n u),\\
a_\phi &= (\gamma/\omega) 2 \pi n.
}
Thus, in agreement with our expectation, the deformation function correctly maps between the two stationary worldline paths.

\subsection{Sinusoidal background}
\label{subsec:sinus}
Another solvable example is given when we assume a sinusoidal background for the deformed model of the form
$\mathbf{E} = E \cos(\omega t) \hat x_3$ \cite{Dunne:2005sx}. The corresponding potential function reads
\eqn{
\tilde F_\phi (\phi) = \sinh(\omega \phi).
\label{eq:Ftilde-sinus}
}
As before, inserting \eqref{eq:Ftilde-stat} and \eqref{eq:Ftilde-sinus} into \eqref{eq:mod-def}, we find
\eqn{
f(\phi) =
\frac{\gamma  \tan \left(\frac{i a_{\chi }}{\gamma  a_{\phi }} \mathbf F\left(i \phi \omega \left|-\frac{1}{\gamma ^2}\right.\right) \right)}{\omega  \sqrt{1+\tan ^2\left(\frac{i a_{\chi }}{\gamma  a_{\phi }} \mathbf F\left(i \phi \omega \left|-\frac{1}{\gamma ^2}\right.\right) \right)}}
}
where
$\mathbf F$ is the incomplete elliptic integral of the first kind \cite{abramowitz}.
The inverse deformation function is given by
\eqn{
f^{-1}(\chi) =
\frac{i \gamma}{\omega} \mathbf{am}\left(\frac{\gamma a_\phi}{a_\chi}  \tanh ^{-1}\left(\frac{\omega \chi }{\sqrt{(\omega \chi)^2-\gamma ^2}}\right) \left|-\frac{1}{\gamma ^2} \right. \right)
}
with $\mathbf{am}$ denoting the Jacobi amplitude \cite{abramowitz}.
Next, inserting the expressions from \eqref{eq:circular-wi} into $f^{-1}$
shows that
\eqn{
\phi &= \frac{i \gamma}{\omega} \mathbf{am}\left(\frac{2 \gamma  \tan ^{-1}\left( \frac{\sin (2 n \pi  u)}{\sqrt{\cos ^2(2 n \pi  u)}}\right) \mathbf K\left(\frac{\gamma ^2}{1 + \gamma ^2}\right)}{i\pi  \sqrt{1 + \gamma ^2}}\left|-\frac{1}{\gamma ^2} \right. \right),\nonumber\\
a_\phi &= \frac{4 \gamma  n \mathbf K\left(\frac{\gamma ^2}{1 + \gamma ^2}\right)}{\omega \sqrt{1 + \gamma ^2} }
}
where $\mathbf K$ is the complete elliptic integral of the first kind, see \cite{abramowitz}.
As can be easily checked, the solution $\phi$ and invariant $a_\phi$ coincide with the findings in \cite{Dunne:2005sx}. This confirms that the deformation function based on \eqref{eq:mod-def} perfectly maps between the two distinct instanton paths.

\subsection{Pulsed background}
\label{subsec:sauter}
Another model that can be exactly solved depends on a pulsed background of Sauter type with the form
$\mathbf{E} = E \mathrm{sech}^2(\omega t) \hat x_3$ \cite{Dunne:2005sx}. The potential function is given by
\eqn{
\tilde F_\phi (\phi) = \tan(\omega \phi).
\label{eq:Ftilde-sauter}
}
As before, inserting \eqref{eq:Ftilde-stat} and \eqref{eq:Ftilde-sauter} into \eqref{eq:mod-def}, we find
\eqn{
f(\phi) =
\frac{\gamma}{\omega} \sin \left( \frac{a_{\chi } \tan ^{-1}\left(\frac{\sqrt{2} \sqrt{\gamma ^2+1} \sin (\phi \omega )}{\sqrt{\gamma ^2+\left(\gamma ^2+1\right) \cos (2 \phi \omega )-1}}\right)}{\sqrt{\gamma ^2+1} a_{\phi }}\right),
}
again given in units of $[m/E]$.
Conversely, the inverse function now comes with both a positive and a negative sign in front given by
\eqn{
&f^{-1}(\chi) =
\pm \frac{\gamma}{\omega }\\
&\times
\cos ^{-1}\left(\frac{\sqrt{\tan ^2\left(\frac{\sqrt{\gamma ^2+1} a_{\phi } \sin ^{-1}\left(\frac{\chi \omega }{\gamma }\right)}{a_{\chi }}\right)+\gamma ^2+1}}{\sqrt{\gamma ^2+1} \sqrt{\tan ^2\left(\frac{\sqrt{\gamma ^2+1} a_{\phi } \sin ^{-1}\left(\frac{\chi \omega }{\gamma }\right)}{a_{\chi }}\right)+1}}\right).\nonumber
}
Plugging the expressions from \eqref{eq:circular-wi} into $f^{-1}$, we write down the final result
\eqnsplit{
\phi &=
\text{sgn}(\sin (2 \pi  n u))\\
&\times \frac{\gamma}{\omega} \cos ^{-1}\left(\frac{\sqrt{\gamma ^2+ \tan ^2(2 \pi  n u) +1}}{\sqrt{\gamma ^2+1} \sqrt{\tan ^2(2 \pi  n u) +1}}\right),\\
a_\phi &= \frac{2 \pi  \gamma  n}{\omega \sqrt{\gamma ^2+1} }.
}
Here, we have chosen the positive sign for $n u \in [0,1/2]$ and the negative one for $n u \in [1/2,1]$ to guarantee the required periodicity of the stationary solution. It can be easily verified that $\phi$ and $a_\phi$ perfectly coincide with the exact solution in \cite{Dunne:2005sx}. Thus, as in the previous example,
the deformation function perfectly maps between the two distinct worldline instantons. 

It should be noticed that the present background is somewhat more interesting, since the smooth deformation between the underlying potential functions, i.e. \eqref{eq:Ftilde-stat} $\rightarrow$ \eqref{eq:Ftilde-sauter}, is not so obvious (due to the presence of poles) as for  \eqref{eq:Ftilde-stat} $\rightarrow$ \eqref{eq:Ftilde-sinus}. This clearly shows that even in the presence of a more complicated background, the function $f^{-1}$ is capable to map the original static instantons to the new stationary worldline paths of the deformed model.

It is important mentioning that in certain cases worldline instantons can curve around branch points or cuts, see e.g. \cite{Ilderton:2015qda}.
In the present study, we have assumed that $f$ is invertible. This is required due to the proposed prescription in \eqref{eq:mod-def} which leads to a deformed model with exact solution $\phi = f^{-1}(\chi)$. 

On the other hand, for backgrounds which give rise to stationary solutions bending around branch points/cuts one would not necessarily end up with stationary equations of the form \eqref{eq:phi-wi}. 

Conversely, when the general form of the background is more complicated compared to our assumption in \eqref{eq:A3}, the function $f$ will be determined by some other equation different from \eqref{eq:mod-def}.
To find out whether and how the deformation idea can be applied to more complicated classes of backgrounds remains an open issue.

\section{Summary and outlook}
\label{Sec:summ}
We have proposed a novel technique permitting the construction of exactly solvable models coupled to external backgrounds whose stationary worldline paths (worldline instantons) determine the Schwinger pair creation rate.

Motivated by the found similarities with earlier ideas on defect solutions in quantum field theories,
we have proposed a deformation function which maps
between the distinct stationary worldline solutions of two separate models.

Starting with a given exactly solvable model, a sequence of iterative deformations generates
a chain of (infinitely many) new exactly solvable models.
The proposed technique may have many advantageous for exact analytical studies of nonstatic backgrounds
in the Schwinger effect.

In the present work, we have focused on purely time-dependent electric backgrounds and confirmed the validity of our results in well-known solvable cases. Further solvable models will be constructed in a forthcoming publication. A potential extension may be the analogous treatment of spatially inhomogeneous backgrounds. 

Moreover, our results may allow an examination of the quantum fluctuation prefactor in more complicated backgrounds.
The proposed technique may also help to better understand the mechanism of dynamical assistance. We plan to investigate these subjects elsewhere.

Apart from these open problems, it remains an interesting question whether an appropriate deformation prescription can be obtained for more complicated backgrounds which go beyond the general form assumed in this paper.

\acknowledgements
I would like to thank Gudi Moortgat-Pick for helpful suggestions. I am grateful to Jim Talbert for carefully reading this work and James Gordon for helpful comments. I gratefully acknowledge the support of the SFB 676 \textit{Particles, Strings and the Early Universe}.

\bibliography{mainbib}
\end{document}